\documentclass[preprint,12pt]{revtex4}
\usepackage{graphicx}
\begin{document}

\title{Inferring Fundamental Value and Crash Nonlinearity from Bubble
  Calibration}

\author{Wanfeng Yan $\dag$, Ryan Woodard $\dag$ and Didier Sornette $\dag \ddag$}
\email{wyan@ehtz.ch, rwoodard@ethz.ch, dsornette@ethz.ch}
\thanks{\\~Corresponding author: Didier Sornette.}
\address{\normalfont{$\dag$ Chair of Entrepreneurial Risks\\
Department of Management, Technology and Economics\\ ETH Z\"{u}rich,
CH-8001 Z\"{u}rich, Switzerland \vspace{12pt}\\ $\ddag$ Swiss
Finance Institute, c/o University of Geneva\\ 40 blvd. Du Pont
dArve, CH 1211 Geneva 4, Switzerland}\\\vspace{12pt}}

\begin{abstract}
  Identifying unambiguously the presence of a bubble in an asset price remains
  an unsolved problem in standard econometric and financial economic
  approaches.  A large part of the problem is that the fundamental value of an
  asset is, in general, not directly observable and it is poorly constrained to
  calculate.  Further, it is not possible to distinguish between an
  exponentially growing fundamental price and an exponentially growing bubble
  price.

  In this paper, we present a series of new models based on the
  Johansen-Ledoit-Sornette (JLS) model, which is a flexible tool to detect
  bubbles and predict changes of regime in financial markets. Our new models
  identify the fundamental value of an asset price and a crash nonlinearity
  from a bubble calibration. In addition to forecasting the time of
  the end of a bubble, the new models can also estimate the fundamental value and the crash
  nonlinearity, meaning that identifying the presence of a bubble is enabled by
  these models.  Besides, the crash nonlinearity obtained in the new models
  presents a new approach to possibly identify the dynamics of a crash after a
  bubble.

  We test the models using data from three historical bubbles ending in crashes from different
  markets. They are: the Hong Kong Hang Seng index 1997 crash, the S\&P 500
  index 1987 crash (black Monday) and the Shanghai Composite index 2009
  crash. All results suggest that the new models perform very well in
  describing bubbles, forecasting their ending times and estimating fundamental
  value and the crash nonlinearity.

  The performance of the new models is tested under both the Gaussian residual
  assumption and non-Gaussian residual assumption.  Under the Gaussian residual
  assumption, nested hypotheses with the Wilks statistics are used and the
  p-values suggest that models with more parameters are necessary. Under
  non-Gaussian residual assumption, we use a bootstrap method to get type I and
  II errors of the hypotheses. All tests confirm that the generalized JLS
  models provide useful improvements over the standard JLS model.

\end{abstract}

 \maketitle \clearpage

\section{Introduction}

Financial bubbles are generally defined as transient upward accelerations of
price above a fundamental value \citep{g97,k00,s03}.  Fundamental value
reflects the intrinsic value (and is sometimes called this) of the asset
itself. It is ordinarily calculated by summing the future incomes generated by
the asset, which are discounted to the present. However, as the future income flow
is uncertain and not known in advance, and since the interest rates that should be used to discount
future cash flows are bound to change in ways not yet known at the time
of the calculation,  the
fundamental value of the asset is usually hard to estimate. In this sense,
identifying unambiguously the presence of a bubble remains an unsolved problem
in standard econometric and financial economic approaches \citep{g08,ls02}.

The Johansen-Ledoit-Sornette (JLS) model
\citep{jsl99,jls00,js99} provides a
flexible framework to detect bubbles and predict changes of regime in the price time
series of a financial asset. It combines (i) the economic theory of rational
expectation bubbles, (ii) behavioral finance on imitation and herding of
investors and traders and (iii) the mathematical and statistical physics of
bifurcations and phase transitions. The model considers the
faster-than-exponential (power law with finite-time singularity) increase in
asset prices decorated by accelerating oscillations as the main diagnostic of
bubbles. It embodies a positive feedback loop of higher return anticipations
competing with negative feedback spirals of crash expectations. Our group has
made many successful predictions using JLS model, such as the 2006 - 2008 oil
bubble \citep{swz09}, the Chinese index bubble in 2009 \citep{jzswbc10}, real
estate market in Las Vegas \citep{zs08}, South African stock market bubble
\citep{zs06} and US Repos market \citep{yws102}.  We also have recently
developed new methods based on this model for forecasting rebounds of the stock
market rather than crashes \citep{yws10}.

In this paper, we generalize the standard JLS model by inferring fundamental
value of the stock and crash nonlinearity from bubble calibration. The new
models can not only detect the crash time but also estimate the fundamental
value and the crash nonlinearity. This means that our new model has the ability
to identify the presence of a bubble, thereby addressing the problem stated at
the beginning of this paper. With the estimated fundamental value, another
famous unsolved problem becomes easier: distinguishing between an exponentially
growing fundamental price and an exponentially growing bubble
price. Furthermore, the new models can also detect the dynamics of crash after
the bubble by specifying how the price evolves towards the fundamental value
during the crash.

We test the models using data from three historical bubbles from different
markets that ended in significant crashes.
 They are: the Hong Kong Hang Seng index 1997 crash, the S\&P 500 index
1987 crash (black Monday) and the Shanghai Composite index 2009 crash. All
results suggest that the new models perform very well in describing bubbles,
forecasting their ending times and estimating fundamental value and the crash
nonlinearity.

The performance of the new models is tested under both the Gaussian residual
assumption and non-Gaussian residual assumption.  Under the Gaussian residual
assumption, nested hypotheses with the Wilks statistics are used and the
p-values suggest that models with more parameters are necessary. Under
non-Gaussian residual assumption, we use a bootstrap method to get type I and
II errors of the hypotheses. All tests confirm that the generalized JLS models
provide useful improvements over the standard JLS model.

The paper is constructed as follows. In Section \ref{sec:jls}, we introduce the
standard JLS model and our new generalized JLS models. We then analyze three
historical bubbles with the new models in Section \ref{sec:case}.  In Section
\ref{sec:stats}, we compare the generalized models statistically to confirm
that these new models provide useful improvements over the standard JLS model.
We conclude in Section \ref{sec:con}.

\section{JLS models}
\label{sec:jls}
\subsection{Standard JLS model}

In the JLS model \citep{jsl99,jls00,js99}, the dynamics of a given
asset is described as
\begin{equation}
  \frac{dp}{p} = \mu(t)dt + \sigma(t)dW - \kappa dj,\label{eq:dynamic}
\end{equation}
where $p$ is the asset price, $\mu$ is the drift (or trend) and $dW$
is the increment of a standard Wiener process (with zero mean and unit
variance). The term $dj$ represents a discontinuous jump such that
$j = 0$ before the crash and $j = 1$ after the crash occurs. The
loss amplitude associated with the occurrence of a crash is
determined by the parameter $\kappa$. Each successive crash
corresponds to a jump of $j$ by one unit. The dynamics of the jumps
is governed by a crash hazard rate $h(t)$. Since $h(t) dt$ is the
probability that the crash occurs between $t$ and $t+dt$ conditional
on the fact that it has not yet happened, we have $ E_t[dj]  = 1
\times h(t) dt + 0 \times (1- h(t) dt)$ and therefore
\begin{equation}
  E_t[dj] = h(t)dt.
\end{equation}
Under the assumption of the JLS model, noise traders exhibit collective herding
behaviors that may destabilize the market. The JLS model assumes that the
aggregate effect of noise and fundamental traders can be accounted for by the
following dynamics of the crash hazard rate
\begin{equation}
  h(t) = B'(t_c-t)^{m-1}+C'(t_c-t)^{m-1}\cos(\omega\ln (t_c-t) -\phi')~.
  \label{eq:hazard}
\end{equation}
If the exponent $m<1$, the crash hazard may diverge as $t$
approaches a critical time $t_c$, corresponding to the end of the
bubble. The second term in the r.h.s. of (\ref{eq:hazard}) takes
into account the existence of a possible hierarchical cascade of
panic acceleration punctuating the course of the bubble, resulting
either from a preexisting hierarchy in noise trader sizes
\citep{zshd05} and/or the interplay between market price impact
inertia and nonlinear fundamental value investing \citep{is02}.

The no-arbitrage condition reads ${\rm E}_t[dp]=0$, which leads to $\mu(t) =
\kappa h(t)$. Taking the expectation of (\ref{eq:dynamic}) with the condition
that no crash has yet occurred gives $dp/p = \mu(t) dt = \kappa h(t) dt$. Using
the crash hazard rate defined in (\ref{eq:hazard}) and integrating yields the
so-called log-periodic power law (LPPL) equation for the price:
\begin{equation}
  \ln p(t) = {\cal F}_{LPPL}(t)~,
\label{eq:lppl}
\end{equation}
where
\begin{equation}
    {\cal F}_{LPPL}(t) = A + B(t_c-t)^m + C(t_c-t)^m\cos(\omega \ln (t_c-t) - \phi)~,
    \label{LPPLforgen}
\end{equation}
$B = - \kappa B' /m$ and $C = - \kappa C' / \sqrt{m^2+\omega^2}$.
Note that this expression (\ref{eq:lppl}) with (\ref{LPPLforgen})
describes the average price dynamics only up to the end of the
bubble. The JLS model does not specify what happens beyond $t_c$.
This critical $t_c$ is the termination of the bubble regime and the
transition time to another regime. For $0<m<1$, the crash hazard rate
accelerates up to $t_c$ but its integral up to $t$, which controls
the total probability for a crash to occur up to $t$, remains finite
and less than $1$ for all times $t \leq t_c$. It is this property
that makes rational for investors to remain invested knowing that a
bubble is developing and that a crash is looming \citep{jls00,
jsl99}. Indeed, there is still a finite probability that no crash
will occur during the lifetime of the bubble, including its end. The excess return
$\mu(t) = \kappa h(t)$ is the remuneration that investors require to
remain invested in the bubbly asset, which is exposed to a crash
risk. The condition that the price remains finite at all time,
including $t_c$, requires that $m \geq 0$.

Within the JLS framework, a bubble is identified when the crash
hazard rate accelerates. According to (\ref{eq:hazard}), such
accelerates occur when $m<1$ and $B'>0$, hence $B<0$ since $m \geq
0$ by the condition that the price remains finite. We thus have a
first condition for a bubble to occur:
\begin{equation}
  0 < m < 1~.
\label{eq:m}
\end{equation}
This condition is the mathematical embodiment of our definition of a
financial bubble, characterized by a faster-than-exponential growth
as time approaches the critical time $t_c$. Indeed, it is
straightforward to verify that the first-order and higher-order
derivatives of the log-price diverge at $t_c$, in contrast with
their finiteness for the standard exponential price model.

By definition, the crash rate should be non-negative. This imposes
\citep{bm03}
\begin{equation}
 b \equiv -Bm - |C|\sqrt{m^2+\omega^2}  \geq 0~.
  \label{eq:bg0}
\end{equation}

\subsection{Modified JLS models}

In an effort to study the fundamental price, we modify and generalize the JLS
model as follows. We now write the price dynamics of an asset as
\begin{equation}
  dp = \mu(t)pdt + \sigma(t)pdW - \kappa (p-p_1)^\gamma dj,\label{eq:dyna2tgmic}
\end{equation}
where the first two items of the right hand side define the standard geometrical
Brownian motion and the third term is the jump.

When the crash occurs at some time $t^*$ (implying $\int_{t^{*-}}^{t^{*+}}
dj=1$), the price drops abruptly by an amplitude $\kappa (p(t^*)-p_1)^\gamma$.

The motivations and the interpretation of the three parameters $p_1, \kappa$
and $\gamma$ are as follows.
\begin{itemize}
\item For $\kappa=\gamma=1$, the price drops from $p(t^{*-})$ to $p(t^{*+})=p_1$,
i.e., the price changes from its value just before the crash to a
fixed well-defined valuation $p_1$. In the spirit of Fama's analysis
of the 19 October 1987 crash \citep{bffmrt89}, if one interprets
the asset price after the crash as the ``right'' price, i.e., the
price discovery towards rational equilibrium without mispricing, the
crash is nothing but an efficient assessment by investors of the
``true'' or fundamental value, once the panic has ended. Hence,
$p_1$ can be interpreted as the fundamental price which is
discovered during the crash dynamics.

\item Then, $\kappa$ can be thought of as a measure of market efficiency, that is,
$1-\kappa$ is the relative inaccuracy of the discovery of the
fundamental price by the market.  If, say, $\kappa=0.5$, this means
that the price has dropped by only half of its bubble component, and
remains over-valued compared with its fundamental component.

\item When different from $1$, the exponent $\gamma$ can be interpreted as
  embodying a nonlinear (i) over-reaction for small variations and
  under-reaction for large deviations ($0<\gamma<1$) or (ii) under-reaction for
  small variations and over-reaction for large deviations ($\gamma >1$) from
  the fundamental value.
\end{itemize}

Since $p_1$ is a fixed parameter, the generalized JLS model implies
that we should measure the price dynamics in the frame moving with
the fundamental price. In other words, $p_1$ is the fundamental
price at the beginning $t_1$ of the time period over which the
bubble develops. In order to compare in a consistent way the
realized price to this fixed parameter, it is necessary to discount
the asset price continuously by the rate of return of the
fundamental price. If $p_{\rm obs}(t)$ denotes the empirical price
observed at time $t$, this means that the price $p(t)$ that enters
in expression (\ref{eq:dyna2tgmic}) is defined by
\begin{equation}
    p(t) = p_{\rm obs}(t)  \prod_{s=t_1+1}^t \frac{1}{(1+r_f(s))^\frac{1}{365}}~,
    \label{discpriceeq}
\end{equation}
where $r_f(s)$ is the annualized growth (risk free) rate of the
fundamental price. In our empirical analysis, we will take for
$r_f(s)$ the annualized US 3-month treasury bill rate.

Applying again the no-arbitrage condition ${\rm E}_t[dp]=0$ to
expression (\ref{eq:dyna2tgmic}) leads to
\begin{equation}
\mu(t)p = \kappa (p-p_1)^\gamma h(t)~.
\end{equation}
Conditional on the absence of a crash, the dynamics of the expected
price obeys the equation
\begin{equation}
  dp = \mu(t)p dt  =  \kappa (p-p_1)^\gamma h(t) dt~,
  \label{eq:dp}
\end{equation}
and the fundamental price must obey the condition $p_1 < \min p(t)$.
For $\gamma = 1$, the solution of equation (\ref{eq:dp}) generalizes
(\ref{eq:lppl}) into
\begin{equation}
  \ln [p(t)-p_1] = {\cal F}_{LPPL}(t) ~,
\label{eq:lpplwrt2tr}
\end{equation}
where ${\cal F}_{LPPL}(t)$ is again given by expression
(\ref{LPPLforgen}). For $\gamma \in (0, 1)$, the solution is
\begin{equation}
    (p-p_1)^{1-\gamma} = {\cal F}_{LPPL}(t)~,
\label{eq:lpplwrt2trgamma}
\end{equation}
where again ${\cal F}_{LPPL}(t)$ is given by expression
(\ref{LPPLforgen}). We do not consider the case $\gamma>1$ which
would give an economically non-sensible behavior, namely the price
diverges in finite time before the crash hazard rate itself
diverges.

In summary, we shall consider four models $M_0$, $M_1$, $M_2$ and $M_3$, where
some are nested in others.  The goal will be to then apply statistical tests to
the models to determine which are sufficient or not and which are necessary or
not.  In the following models, ${\cal F}_{LPPL}(t)$ below is given by
expression (\ref{LPPLforgen}).
\begin{enumerate}
\item[0.] Original JLS model $M_0$: $p_1 = 0, \gamma = 1$ (with $\kappa < 1$):
\begin{equation}
  p_{M_0}(t) = \exp({\cal F}_{LPPL}(t)) ~.
\label{eq:m0}
\end{equation}

\item[1.] $M_1$: $p_1 \neq 0, \gamma = 1$:
\begin{equation}
  p_{M_1}(t) = p_1 + \exp({\cal F}_{LPPL}(t))~.
  \label{eq:m1}
\end{equation}
$M_1$ includes $M_0$ as a special case. In other words, $M_0$ is
nested in $M_1$.

\item[2.] $M_2$: $p_1 = 0, \gamma \in (0, 1]$:
\begin{equation}
    p_{M_2}(t) = \left\{ \begin{array}{ll}
        \left({\cal F}_{LPPL}(t)\right)^{\frac{1}{1-\gamma}}, &\gamma \in (0,1)~,\\
        \exp\left({\cal F}_{LPPL}(t)\right), &\gamma = 1~.
\end{array}
\right. \label{eq:m2}
\end{equation}
Since $M_2$ includes $M_0$ as a special case, $M_0$ is also nested
in $M_2$.

\item[3.] $M_3$: $p_1 \neq 0, \gamma \in (0, 1]$:
\begin{equation}
\begin{array}{ll}
    p_{M_3}(t) =
    \left\{\begin{array}{rcl}
        p_1 + \left({\cal F}_{LPPL}(t)\right)^{\frac{1}{1-\gamma}}, &\gamma \in (0,1)~,\\
        p_1 + \exp\left({\cal F}_{LPPL}(t)\right), &\gamma = 1~.
        \end{array}
\right.
\end{array}
\label{eq:m3}
\end{equation}
$M_3$ includes all previous models, $M_0, M_1$ and $M_2$ as special
cases, so that $M_0, M_1$ and $M_2$ are all nested in $M_3$.
\end{enumerate}

\section{Calibration and results on three historical bubbles}
\label{sec:case}
\subsection{Calibration method of the models}

Given an observed asset time series of prices $\{p_{\rm obs}(t)\}$, we first
transform it into a price time series of discounted prices $\{p(t)\}$ by using
expression (\ref{discpriceeq}).  We next determine the three parameters $A, B$
and $C$ in expression (\ref{LPPLforgen}) for each model as a function of the
other parameters, by solving analytically the system of three linear equations
obtained by minimizing the square of deviations:
\begin{itemize}
\item $\ln[p(t)]- {\cal F}_{LPPL}(t)$ for $M_0$,
\item $\ln[p(t)-p_1]- {\cal F}_{LPPL}(t)$ for $M_1$,
\item $[p(t)]^{1-\gamma}- {\cal F}_{LPPL}(t)$ for $M_2$ and
\item $[p(t)-p_1]^{1-\gamma}- {\cal F}_{LPPL}(t)$ for $M_3$.
\end{itemize}
We then determine the other parameters for each model using a Taboo search (to
find initial parameter estimates) coupled with a Levenberg-Macquardt
algorithm. We constrain the values of plausible parameters as follows:
\begin{enumerate}
\item the fundamental price $p_1$ should be larger than $0.2 p_{\rm min}$,
  where $p_{\rm min}:={\rm Min}[p(t)]$ over the fitting time interval.
\item The fit parameters $t_c$, $m$, $p_1$ and $\gamma$ should not be on the
  boundary of the search intervals. They should deviate from these boundaries
  by at least 1\% in relative amplitude.
\item Among all the fits satisfying the above two conditions, the one with the
  smallest sum of normalized residuals is selected.  The cost function we use
  here is the sum of squares of the relative discounted price differences
\begin{equation}
    R(t) = \frac{p(t) - p_M(t)}{p_M(t)}~,
    \label{eq:r}
\end{equation}
where $p_M(t)$ stands for one of the expressions
(\ref{eq:m0}-\ref{eq:m3}).

\end{enumerate}
The critical time $t_c$ corresponding to the end of the bubble is
searched in $[t_2; t_2 + 0.4 (t_2-t_1)]$, where the time window of
analysis is $[t_1; t_2]$. The exponent $m$ is constrained in
$[10^{-5}; 1-10^{-5}]$. The log-angular frequency $\omega$ is
searched in $[0.01; 40]$. The phase $\phi$ can take values in $[0,
2\pi-10^{-5}]$. The fundamental price $p_1$ is in $[0.01; 0.99
p_{\rm min}]$ and then restricted by condition (i) above.

\subsection{Results}
\label{sec:result}

We calibrate models $M_0-M_3$ to three well-documented bubbles,
which ended in large crashes:
\begin{itemize}
\item Hong Kong Hang Seng index (HSI) ($t_1 =$ Feb. 1, 1995, $t_2 =$ March 13,
  1997),
\item S\&P 500 index (GSPC) ($t_1 =$ Sept. 1, 1986, $t_2 =$ Aug. 26, 1987),
\item Shanghai Composite index (SSEC) ($t_1 =$ Oct. 24, 2008, $t_2 =$ July 10,
  2009.
\end{itemize}

\subsubsection{Presentation and discussion}

\begin{figure}
\centering
\includegraphics[width=\textwidth]{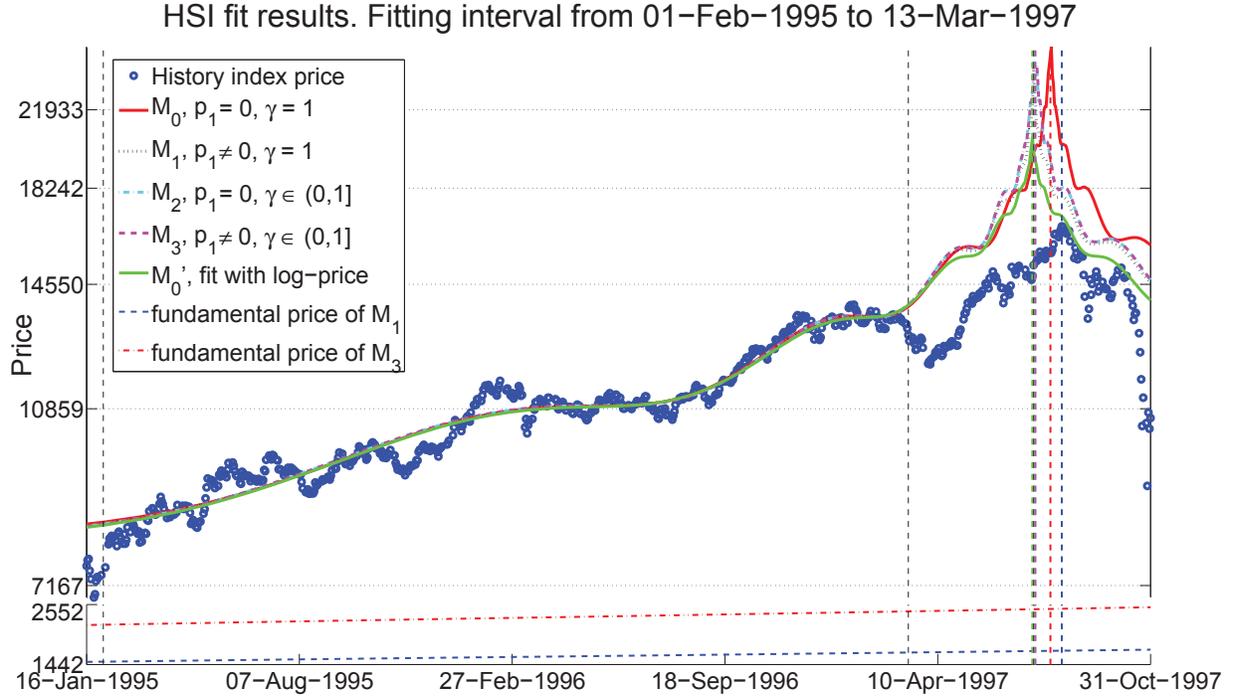}
\caption{Calibration of the different models to the Hong Kong Hang
Seng Index. The fit interval is shown with vertical black dashed
lines. The fitted critical time $t_c$ when the crash is most
probable according the modified JLS models are marked by vertical
dashed lines with the same color as the corresponding fits with each
model. The historical close prices are shown as blue empty circles.
The fundamental price for $M_1$ and $M_3$ are also shown as the
almost horizontal dashed lines (beware of the break in the vertical
scales for low values).} \label{fig:hsi}
\end{figure}

\begin{figure}
\centering
\includegraphics[width=\textwidth]{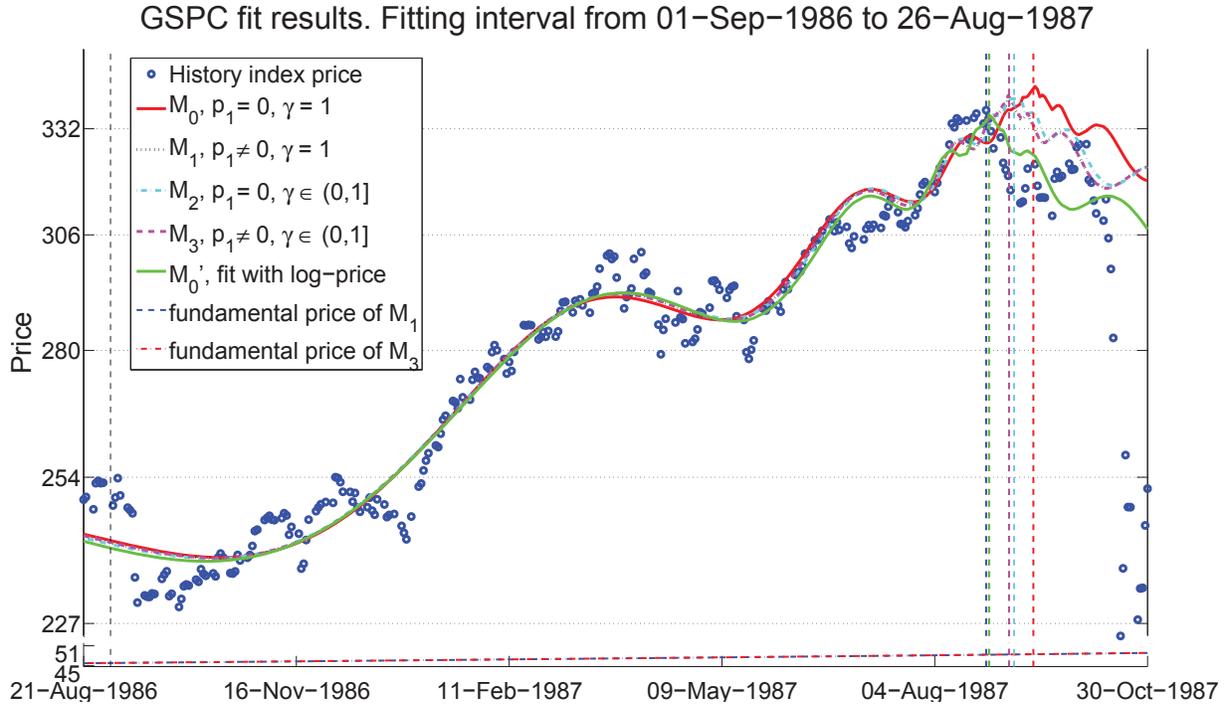}
\caption{Same as figure \protect\ref{fig:hsi} for the S \& P 500
Index. } \label{fig:gspc}
\end{figure}

\begin{figure}
\centering
\includegraphics[width=\textwidth]{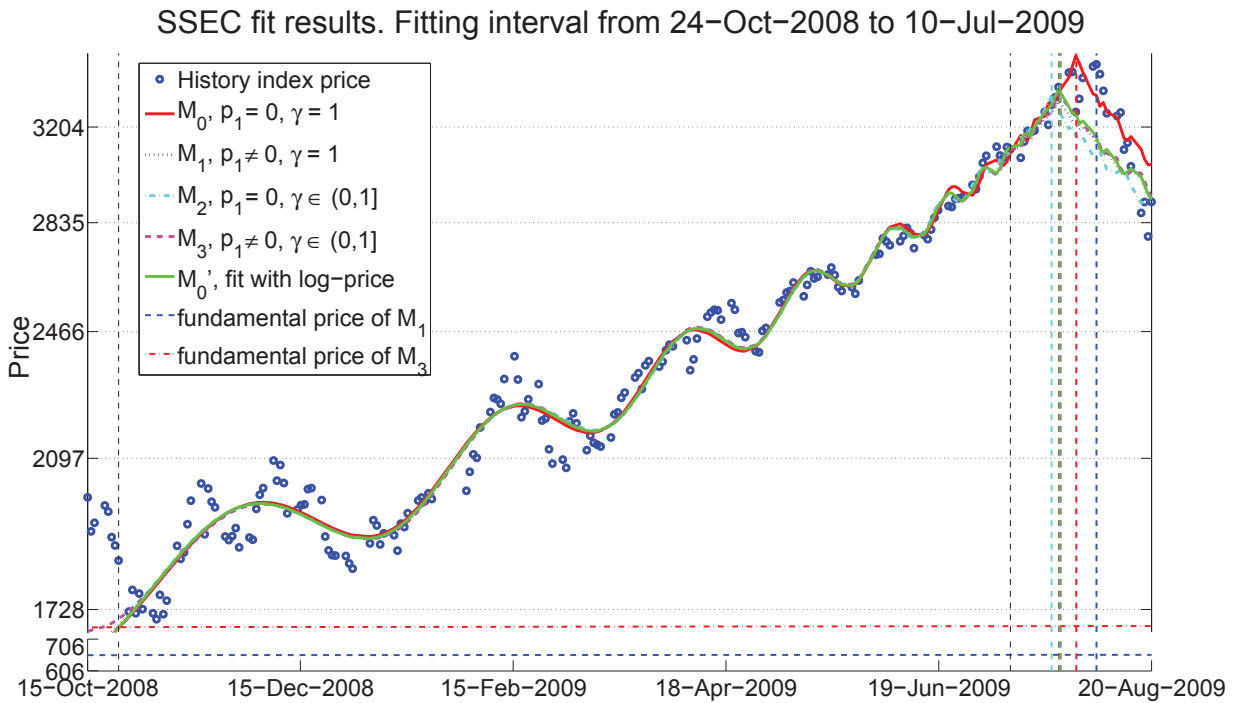}
\caption{Same as figure \protect\ref{fig:hsi} for the Shanghai
Composite Index.} \label{fig:ssec}
\end{figure}

The results are shown in Figs.~\ref{fig:hsi} - \ref{fig:ssec} and the
corresponding parameters are given in Tables~\ref{tb:hsiparm} -
\ref{tb:ssecparm}. Visually, all models seem to perform similarly, with the
determined critical times $t_c$ close to the true time of the crash. We note
that the parameters $p_1$ and $\gamma$ in $M_1, M_2$ and $M_3$ depart
significantly from their reference values $p_1=0$ and $\gamma=1$ characterizing
model $M_0$.

Model $M'_0$ corresponds to model $M_0$ with a slightly different
cost-function. Instead of minimizing the sum of the squares of terms given by
(\ref{eq:r}), for $t$ going from $t_1$ to $t_2$, the parameters of $M'_0$ are
those of model $M_0$ obtained by minimizing the sum of the squares of the
difference $\ln [p_{M_0}(t)]- {\cal F}_{LPPL}(t)$. Since $\ln y -\ln x =
(y-x)/x + {\cal O}[(y-x)/x]^2$, the two methods should give similar results and
the results summarized in tables \ref{tb:hsiparm}-\ref{tb:ssecparm} confirm
this expectation.

Results of detailed statistical comparisons between the four models are shown
below. Tables \ref{tb:hsiparm}-\ref{tb:ssecparm} suggest that the five models
perform almost equivalently in their ability to fit the price accelerations and
to determine the time $t_c$ of the peak of the bubbles. One can note a
remarkable stability and consistency of the estimators for the two crucial
parameters, the exponent $m$ and the angular log-frequency $\omega$. However,
models $M_1$ and $M_3$ provide an interesting estimation of the size of the
bubble, which appears stable with respect to these two specifications: at the
beginning of the calibration interval, for the Hong Kong bubble, models $M_1$
and $M_3$ estimate that the bubble component might have been already accounting
for $71\%$ to $80\%$ of the observed price. At the end of the bubble, the
bubble component is between $85\%$ to $90\%$ of the observed price. Similar
values are found for the two other case studies.  An exception is for the
Shanghai Composite index bubble, for which model $M_3$ suggests that the
fundamental price was $92\%$ of the observed price at the beginning of the
calibrating interval and about half of the observed price at its peak.

The models provide a method to measure the amplitude of the crash that follows
the bubble peak. Consider two types of drawdown after the peak: (i) $DD_{\rm
  [2months]}$ is the two-months drop measured from the peak; (ii) $DD_{\rm
  max}$ is the peak-to-valley drawdown from the peak to the minimum of the
asset price after the crash. We calculate the magnitude of the crash
compared to the over-valued prices as follows. The ratio between the crash magnitude and
over-valued prices is estimated as:
\begin{equation}
RC_i = \frac{DD_i}{p_{\rm obs}(t_p) - p_1  \prod_{s=t_1+1}^t
(1+r_f(s))^\frac{1}{365}}~~~~~ i \in \{{\rm[2months]}, {\rm max}\} .
\label{ththntnbw}
\end{equation}
During the crashes, the hazard rate in Eq. ~\ref{eq:dp}) should be $1$.  Then
comparing the definition of $RC$ and Eq.~(\ref{eq:dp}), one can easily find that
$\kappa = RC$ for the models whose $\gamma = 1$ ($M_0, M_1, M_0'$). For the
other models, $\kappa$ is different from $RC$.  These values are reported in
tables \ref{tb:hsiparm}-\ref{tb:ssecparm}.

\subsubsection{Consistency test of the calibrations}

According to the specification of \cite{lrs09}, we should verify
that the calibrations discussed above are self-consistent, i.e., the
residuals are stationary. This verification step was proposed by
\cite{lrs09} as a possible solution to the problems identified by
\cite{gn74} and \cite{p86} resulting from the calibration of
non-stationary prices.

In order to check that the normalized residuals are stationary for all the four
models, we use the Phillips-Perron and the Dickey-Fuller unit root tests. The
null hypothesis $H_0$ is that the normalized residuals are not stationary,
i.e. they have a unit root. In order to have reasonable statistics, we consider
time windows of fixed length of 175, 250 or 550 trading days.  We identify
these windows in time series much larger than the $(t_1, t_2)$ intervals used
to identify the bubbles (given at the top of Sec.~\ref{sec:result}). The
interval lengths correspond to the different values of $t_2 - t_1$ for the
respective case studies. We choose overlapping intervals with the start of
neighboring intervals separated by 25 days.  There are 303 windows of size 550
trading days for the HSI from Jan. 1, 1987 to Feb. 25, 2010; 800 windows for
the GSPC index from Feb. 2, 1954 to Feb. 10, 2010 with size of 250 trading
days; 167 windows of SSEC from Aug. 3, 1997 to Jan. 22, 2010 of size of 175
trading days. Note that we choose these dates as the window boundaries because:
(i) the chosen $(t_1, t_2)$ intervals identified at the top of
Sec.~\ref{sec:result} should be one of the windows we get here; (ii) up to the
data collection date (Feb. 26, 2010), we want to get as many windows as we
can. Using the statistical confidence level of 99\%, we determine the fraction
of those windows which reject the Phillips-Perron and the Dickey-Fuller unit
root tests ($H_1$), i.e., which qualify as stationary. The results are
presented in table \ref{tb:stationary}. We conclude that most of the residuals
are found stationary, which support the validity of our calibration procedure.

Previous works have identified the domain of parameters of the
calibration of the JLS model $M_0$ which is the most relevant
(Johansen and Sornette, 2006; Jiang et al., 2010). These conditions,
referred to as the LPPL (log-period power law) conditions, are
\begin{equation}
B > 0;  ~~~~ 0.1 \leq m \leq 0.9; ~~~~ 6 \leq \omega \leq 13; ~~~~
-1 \leq C \leq 1~. \label{hyhy35u6jk4uj}
\end{equation}
Imposing that the calibrations obey these LPPL conditions
(\ref{hyhy35u6jk4uj}), we find in  Table \ref{tb:stlppl} that the
fraction of the above windows analyzed in Table \ref{tb:stationary}
which fulfill the stationary conditions is significantly increased,
augmenting our trust of the quality of the calibration and of the
relevance of  this class of models.

\section{Statistical comparisons of the four generalized JLS models}
\label{sec:stats}

\subsection{Standard Wilks test of nested hypotheses assuming independent and normally distributed residuals}

Let us consider the five pairs of models with nested structure:
$(M_0 \subset M_1)$,  $(M_0 \subset  M_2)$, $(M_1 \subset  M_3)$,
$(M_2 \subset  M_3)$, and $(M_0 \subset  M_3)$.  Let us denote $M_l$
as the model with the smaller number of parameters and $M_h$ that
with the larger number of parameters. For each pair, we use Wilks
test of nested hypotheses in terms of the log-likelihood ratios to
decide between the two hypotheses:
\begin{enumerate}
\item[$H_0$: ] $M_l$ is sufficient and $M_h$ is not necessary.
\item[$H_1$: ] $M_l$ is not sufficient and $M_h$ is needed.
\end{enumerate}
We first present in this subsection the tests assuming that the residuals of
the calibration of the models to the asset price time series are normally and
independently distributed.  In the next subsection, we loosen this restriction.

For each model $M_i$, $i = 0, 1, 2, 3$, let us denote the normalized
residuals defined by expression (\ref{eq:r}) by $R_i(t)$ and assume
that they are i.i.d. Gaussian. For sufficiently large time windows,
and noting $N$ the number of trading days in the fitted time window
$[t_1; t_2]$, the Wilks log-likelihood ratio reads
\begin{equation}
T = 2 \log \frac{L_{h,max}}{L_{l,max}} = 2N \ln
\frac{\sigma_l}{\sigma_h} + \frac{\sum_{t=1}^N R_l^2(t)}{\sigma_l^2}
- \frac{\sum_{t=1}^N R_h^2(t)}{\sigma_h^2}~, \label{ththwfq}
\end{equation}
where $R_l$ and $\sigma_l$ (respectively $R_h$ and $\sigma_h$) are
the residuals and their corresponding standard deviation for $M_l$
(respectively $M_h$).

In the large $N$ limit, and under the above conditions of asymptotic
independence and normality, the $T$-statistics is distributed with a
$\chi_k^2$ distribution with $k$ degrees of freedom, where $k$ is
the difference between the number of parameters in $M_h$ and $M_l$.
We have $k=1$ for the pairs  $(M_0, M_1)$,  $(M_0, M_2)$, $(M_1,
M_3)$, $(M_2, M_3)$, and $k=2$ for $(M_0, M_3)$. The $p$-values
associated with the $T$-statistics given by (\ref{ththwfq}) for each
of the five pairs are reported in Table~\ref{tb:gauss}.  The summary of that
table is:
\begin{itemize}
\item Hong Kong Hang Seng index (HSI) from Feb. 1, 1995 to March 13, 1997:
  Model $M_0$ is never rejected and the standard JLS model is sufficient.

\item S\&P 500 index (GSPC) from Sept. 1, 1986 to Aug. 26, 1987: Model $M_0$ is
  rejected with strong statistical confidence in favor of $M_1$, $M_2$ and
  $M_3$. However, when comparing $M_1$ and $M_2$ to $M_3$, we find that $M_3$
  is not necessary. Therefore, we conclude that the structure of the S\&P 500
  index bubble requires the introduction of either a fundamental price $p_1$ or
  of a nonlinear crash amplitude as a function of mispricing (price for $M_0$
  and $M_2$), but that both ingredients together are not necessary.

\item Shanghai Composite index (SSEC) from Oct. 24, 2008 to July 10, 2009: Only
  $M_3$ improves on $M_0$ at a confidence level of $92.3\%$ that can be
  considered as acceptable, while $M_1$ and $M_2$ are not significantly better
  than $M_0$ for standard confidence levels.  Consistent with $M_3$ being
  rather significantly better than $M_0$, it is also better than $M_1$ and
  $M_2$, which are themselves not significantly improving on $M_0$. There seems
  to exist both a fundamental value component and a nonlinear over-reaction to
  mispricing in the unfolding of this Chinese bubble.
\end{itemize}

\subsection{Comparison between models by bootstrapping
to account for non-normality and dependence between residuals}

Consider a pair of models $(M_l \subset M_h)$. Let us assume that $M_l$ is the
correct generating model of the data. The calibration of $M_l$ to the data
gives a specific set of parameters as well as a specific realization of
residuals. We then use this specification of the model $M_l$ and its residuals
to generate 1000 synthetic time series. A given synthetic time series is the
calibrated $M_l$ time series on which we add residuals obtained by randomly
reshuffling the previously obtained residuals. Thus, the 1000 synthetic time
series differ from each other only by the reshuffling of the residuals. We then
calibrate the two models $M_l$ and $M_h$ on each of these 1000 synthetic time
series and calculate the difference of the sum of the square of residuals of
the fits of these two models.  We thus have a list of 1000 different $d_n$,
$n=1, ..., 1000$.  Comparing with the corresponding difference $d_{\rm fit}$
(between $M_l$ and $M_h$) gives us a realistic estimation of the $p$-value
for the null hypothesis that $M_l$ is the correct generating model of the data.
Specifically, the $p$-value is the fraction among the 1000 $d_n$'s that are
{\it larger} than $d_{\rm fit}$.  For instance, if all values $d_n$ are smaller
than $d_{\rm fit}$, we obtain $p=0$, i.e., it is very improbable that the
difference in quality of fit between $M_l$ and $M_h$ results solely from the
structure of the models and of the residues. We can reject the null and
conclude that $M_h$ is a better necessary model.

The second test we perform starts with the hypothesis that the true generating
process is $M_h$. Thus, the 1000 synthetic time series are now generated by
using model $M_h$ calibrated on the data and its residuals. Then, the $p$-value
for this null is determined as the fraction among the 1000 $d_n$'s that are
{\it smaller} than $d_{\rm fit}$.

Table \ref{tb:nongauss} summarizes the results, which improve on those shown in
Table \ref{tb:gauss} by relaxing the conditions of normality and of
independence between the daily residuals of the calibration. The bootstraps are
performed by reshuffling the residuals of the fit ``every day'' or in blocks of
25 continuous days (``every 25 days''), which is in blocks of 25 continuous
days. The later allows us to keep the dependence structure over 25 days to test
its possible impact on the $p$-values. Reshuffling every day destroys any
dependence in the residuals, while keeping their one-point (possibly
non-Gaussian) statistics.

For HSI, taking into account the dependence structure of the residuals up to 25
days confirm the results already found in Table \ref{tb:gauss} that the
standard JLS model $M_0$ is sufficient to explain the observed financial
bubble. For GSPC, the results also confirm those of the Wilks test in Table
\ref{tb:gauss}, that $M_1$ and $M_2$ improve significantly on $M_0$, while
$M_3$ is not necessary. For SSEC, also in agreement with Table \ref{tb:gauss},
model $M_3$ is found to be the best and to be significant at the $95\%$
confidence level.

Overall, these tests confirm that the generalized JLS models seem to provide
useful improvements over the standard JLS model, both in terms of their
explanatory power and in the extraction of additional information, specifically
the fundamental price $p_1$ and a possible nonlinear dependence of the crash
amplitude as a function of mispricing.

\section{Conclusion}
\label{sec:con}

In this paper, we generalized the JLS model by inferring the
fundamental value and crash nonlinearity from bubble calibration. In
the generalized model, one can not only predict the crash time of a
stock, but also estimate the fundamental value of that stock.
Besides, the crash nonlinearity can also be estimated.

Three historical bubbles from different markets are tested by the
generalized models. All the results suggest that the new models
perform very well in describing bubbles, predicting crash time and
estimating fundamental value and the crash nonlinearity.

The performance of the new models is tested both under the Gaussian
and non-Gaussian residual assumptions. Under the Gaussian residual
assumption, nested hypothesis testing with the Wilks statistics is used and
the p-values suggest models with more parameters are necessary.
Under non-Gaussian residual assumption, we use bootstrap method and
get the type I and II errors of the hypothesis. All those tests
confirm that the generalized JLS models provide useful improvements
over the standard JLS model.

\clearpage
%table 1
\begin{table}
\begin{center}
\begin{tabular}{|l|l|l|l|l|l|l|l|l|l|l|l|}
\hline
HSI&$t_c$&$|t_c-t_p|$&$m$&$\omega$&$\phi$&$\frac{p_f(t_1)}{p(t_1)}$&$\frac{p_f(t_p)}{p(t_p)}$&$\gamma$&$RC_{\rm
[2months]}$&$RC_{\rm max}$&RMS\\\hline
$M_0$&27-Jul-1997&10&0.19&6.97&0.00&-&-&-&0.46&0.62&0.0320\\\hline
$M_1$&11-Jul-1997&26&0.25&6.63&0.78&0.20&0.10&-&0.52&0.69&0.0320\\\hline
$M_2$&12-Jul-1997&25&0.03&6.64&0.87&-&-&0.13&0.46&0.62&0.0319\\\hline
$M_3$&12-Jul-1997&25&0.03&6.65&4.04&0.29&0.15&0.11&0.54&0.73&0.0319\\\hline
$M_0'$&09-Jul-1997&28&0.39&6.53&3.30&-&-&-&0.41&0.55&0.0323\\\hline
\end{tabular}
\caption{Results of the calibration of models $M_0-M_3$ for the Hong
Kong Hang Seng index (HSI) from Feb. 1, 1995 to March 13, 1997.
$t_c$ is the critical time of a given model corresponding to the end
of the bubble and the time at which the crash is the most probable.
$t_1$ is the beginning of the fitting interval. $t_p$ is the time
when the asset value peaks before the crash. The relative amplitude
of the crash following the peak of the bubble is given by $RC_{\rm
[2months]}$ and $RC_{\rm max}$, which
are calculated using expression (\ref{ththntnbw}) from the following
drawdown amplitudes: (i) $DD_{\rm [2months]}$ is
the two-months drop measured from the peak; (ii) $DD_{\rm max}$ is
the peak-to-valley drawdown from the peak to the minimum of the
asset price. RMS is the root mean square of the distances between
historical prices and the model values, i.e., the square root of the
sum of the squares of terms given by (\ref{eq:r}), for $t$ going
from $t_1$ to $t_2$, where $t_2$ is the last date of the time
window used for the analyses. The model denoted $M'_0$ corresponds to model
$M_0$ with a different calibration method, as explained in the
text.} \label{tb:hsiparm}
\end{center}
\end{table}

%table 2
\begin{table}
%%  \centering
    \begin{tabular}{|l|l|l|l|l|l|l|l|l|l|l|l|}
      \hline
      GSPC&$t_c$&$|t_c-t_p|$&$m$&$\omega$&$\phi$&$\frac{p_f(t_1)}{p(t_1)}$&$\frac{p_f(t_p)}{p(t_p)}$&$\gamma$&$RC_{\rm
        [2months]}$&$RC_{\rm max}$&RMS\\\hline
      $M_0$&13-Sep-1987&19&0.70&6.62&0.00&-&-&-&0.34&0.35&0.0196\\\hline
      $M_1$&03-Sep-1987&9&0.68&6.10&0.00&0.18&0.14&-&0.40&0.40&0.0190\\\hline
      $M_2$&05-Sep-1987&11&0.63&6.09&0.00&-&-&0.72&0.34&0.35&0.0191\\\hline
      $M_3$&03-Sep-1987&9&0.64&6.10&0.00&0.18&0.14&0.64&0.40&0.40&0.0190\\\hline
      $M_0'$&26-Aug-1987&1&0.68&5.59&0.14&-&-&-&0.32&0.33&0.0187\\\hline
    \end{tabular}
    \caption{Same as Table \protect\ref{tb:hsiparm} for the S\&P 500
      index (GSPC) from Sept. 1, 1986 to Aug. 26, 1987.}
    \label{tb:gspcparm}
\end{table}

%table 3
\begin{table}
\begin{center}
\begin{tabular}{|l|l|l|l|l|l|l|l|l|l|l|l|}
\hline
SSEC&$t_c$&$|t_c-t_p|$&$m$&$\omega$&$\phi$&$\frac{p_f(t_1)}{p(t_1)}$&$\frac{p_f(t_p)}{p(t_p)}$&$\gamma$&$RC_{\rm
[2months]}$&$RC_{\rm max}$&RMS\\\hline
$M_0$&29-Jul-2009&2&0.63&16.60&0.00&-&-&-&0.23&0.23&0.0258\\\hline
$M_1$&24-Jul-2009&3&0.77&15.86&1.94&0.36&0.19&-&0.29&0.29&0.0256\\\hline
$M_2$&21-Jul-2009&6&0.69&15.52&6.28&-&-&0.99&0.23&0.23&0.0257\\\hline
$M_3$&24-Jul-2009&3&0.65&15.96&2.49&0.92&0.49&0.20&0.45&0.45&0.0254\\\hline
$M_0'$&24-Jul-2009&3&0.68&15.86&5.12&-&-&-&0.23&0.23&0.0256\\\hline
\end{tabular}
\caption{Same as Table \protect\ref{tb:hsiparm} for the Shanghai
Composite index (SSEC) from Oct. 24, 2008 to July 10, 2009.}
\label{tb:ssecparm}
\end{center}
\end{table}

%table 4
\begin{table}
\begin{center}
\begin{tabular}{|l|l|l|l|l|}
\hline\hline Percentage of
stationary&$M_0$&$M_1$&$M_2$&$M_3$\\\hline\hline
\multicolumn{5}{|c|}{303 HSI windows from Jan. 1, 1987 to Feb. 25,
2010, length 550.}\\\hline
Phillips-Perron&96.7\%&98.0\%&96.7\%&97.7\%\\\hline
Dickey-Fuller&96.7\%&98.0\%&96.7\%&97.7\%\\\hline\hline
\multicolumn{5}{|c|}{800 GSPC windows from Feb. 2, 1954 to Feb. 10,
2010, length 250.}\\\hline
Phillips-Perron&90.6\%&91.0\%&91.8\%&91.8\%\\\hline
Dickey-Fuller&90.6\%&91.0\%&91.8\%&91.8\%\\\hline\hline
\multicolumn{5}{|c|}{167 SSEC windows from Aug. 3, 1997 to Jan. 22,
2010, length 175.}\\\hline
Phillips-Perron&96.4\%&97.0\%&96.4\%&97.0\%\\\hline
Dickey-Fuller&96.4\%&97.0\%&96.4\%&97.0\%\\\hline\hline
\end{tabular}
\caption{Percentage of stationary residuals for the Phillips-Perron
and Dickey-Fuller tests. Significance level: 99\%.}
\label{tb:stationary}
\end{center}
\end{table}

%table 5
\begin{table}
\begin{center}
\begin{tabular}{|l|l|l|l|l|}
\hline\hline Percentage of stationary under LPPL
constrains&$M_0$&$M_1$&$M_2$&$M_3$\\\hline\hline
\multicolumn{5}{|c|}{303 HSI windows from Jan. 1, 1987 to Feb. 25,
2010, length 550.}\\\hline
$P_{LPPL}$&0.99\%&0.99\%&2.64\%&1.98\%\\\hline
Phillips-Perron&100\%&100\%&100\%&100\%\\\hline
Dickey-Fuller&100\%&100\%&100\%&100\%\\\hline\hline
\multicolumn{5}{|c|}{800 GSPC windows from Feb. 2, 1954 to Feb. 10,
2010, length 250.}\\\hline
$P_{LPPL}$&4.50\%&6.00\%&4.50\%&5.87\%\\\hline
Phillips-Perron&95.7\%&100\%&97.9\%&100\%\\\hline
Dickey-Fuller&95.7\%&100\%&97.9\%&100\%\\\hline\hline
\multicolumn{5}{|c|}{167 SSEC windows from Aug. 3, 1997 to Jan. 22,
2010, length 175.}\\\hline
$P_{LPPL}$&4.19\%&4.79\%&8.38\%&9.58\%\\\hline
Phillips-Perron&93.8\%&92.9\%&100\%&100\%\\\hline
Dickey-Fuller&93.8\%&92.9\%&100\%&100\%\\\hline\hline
\end{tabular}
\caption{Percentage of stationary residuals, as qualified
by the Phillips-Perron
and Dickey-Fuller tests, which obey the LPPL conditions (\ref{hyhy35u6jk4uj}).
The variable $P_{LPPL}$ gives the
fraction of fits that satisfy the conditions (\ref{hyhy35u6jk4uj}), independently
of whether their residuals are stationary or not.
Significance level: 99\%. } \label{tb:stlppl}
\end{center}
\end{table}

%table 6
\begin{table}
\begin{center}
\begin{tabular}{|l|l|l|l|l|l|}
\hline &$(M_0, M_1)$&$(M_0, M_2)$&$(M_1, M_3)$&$(M_2, M_3)$&$(M_0,
M_3)$\\\hline HSI&0.4710&0.2210&0.3221&0.9626&0.4723\\\hline
GSPC&0.0003&0.0006&0.7930&0.2150&0.0012\\\hline
SSEC&0.1405&0.2494&0.0863&0.0516&0.0775\\\hline
\end{tabular}
\caption{$p$-value of the null hypothesis $H_0$ for pairs of models
$(M_l, M_h)$  that $M_l$ is sufficient and $M_h$ is not necessary,
using  Wilks log-likelihood ratio statistics. Low $p$-value
indicates the improvement of $M_h$ compared to $M_l$ is significant
and $H_0$ is rejected.} \label{tb:gauss}
\end{center}
\end{table}

%table 7
\begin{table}
\begin{center}
\begin{tabular}{|l|l|l|l|l|l|}
\hline\hline &$(M_0, M_1)$&$(M_0, M_2)$&$(M_0, M_3)$&$(M_1,
M_3)$&$(M_2, M_3)$\\\hline\hline \multicolumn{6}{|c|}{HSI shuffle
every day}\\\hline $M_l$ true &0&0&0&0.05&0.75\\\hline $M_h$ true
&0&0&0&0.10&0.60\\\hline\hline \multicolumn{6}{|c|}{HSI shuffle
every 25 days}\\\hline $M_l$ true &0.46&0.20&0.42&0.26&0.76\\\hline
$M_h$ true&0.42&0.12&0.38&0.18&0.70\\\hline\hline
\multicolumn{6}{|c|}{GSPC shuffle every day}\\\hline $M_l$
true&0&0&0&0.35&0.45\\\hline $M_h$
true&0.05&0&0&0.45&0.40\\\hline\hline \multicolumn{6}{|c|}{GSPC
shuffle every 25 days}\\\hline $M_l$
true&0.05&0&0.05&0.40&0.50\\\hline $M_h$
true&0&0&0&0.50&0.45\\\hline\hline \multicolumn{6}{|c|}{SSEC shuffle
every day}\\\hline $M_l$ true&0&0&0&0.05&0.35\\\hline $M_h$
true&0&0.05&0&0.05&0.50\\\hline\hline \multicolumn{6}{|c|}{SSEC
shuffle every 25 days}\\\hline $M_l$
true&0.14&0.08&0.04&0.04&0.38\\\hline $M_h$
true&0.12&0.06&0.06&0.08&0.40\\\hline\hline
\end{tabular}
\caption{$p$-values calculated by bootstraping (see text for
explanation). Low $p$-value indicates the improvement of $M_h$
compared to $M_l$ is significant.} \label{tb:nongauss}
\end{center}
\end{table}

\clearpage
%figure 1

%figure 2

%figure 3

\end{document}